\documentclass[12pt]{article}
\usepackage{float}
\usepackage{epsfig}

\newcommand{\beq}{\begin{equation}}
\newcommand{\eeq}{\end{equation}}

\newcommand{\be}{\begin{equation}}
\newcommand{\ee}{\end{equation}}
\newcommand{\bea}{\begin{eqnarray}}
\newcommand{\eea}{\end{eqnarray}}

\newcommand{\ba}{\begin{array}}   %  
\newcommand{\ea}{\end{array}}     %  
\begin{document}
\baselineskip=17pt

\gdef\journal#1, #2, #3, 1#4#5#6{{#1~}{\bf #2}, #3 (1#4#5#6)}
\gdef\ibid#1, #2, 1#3#4#5{{\bf #1} (1#3#4#5) #2}

\begin{flushright}
July 2002 \\
cond-mat/0207667
\end{flushright}
\sloppy
\vskip 0.7cm

\begin{center}
{\Large The Lowest Landau Level Anyon Equation of State
in the Anti-screening Regime}\\[0.5cm]
{\large  Stefan Mashkevich}\\[0.1cm]
Bogolyubov Institute for Theoretical Physics,
03143 Kiev, Ukraine\footnote{Present address:
Schr\"odinger, 120 West 45th St., New York, NY 10036, USA.
$<$mash@mashke.org$>$}\\[0.2cm]
{\large St\'ephane Ouvry}\footnote{Corresponding author.
$<$ouvry@ipno.in2p3.fr$>$}\\[0.1cm]
Laboratoire de Physique Th\'eorique et Mod\`eles 
Statistiques\footnote{{\it Unit\'e de
Recherche de l'Universit\'e Paris 11 associ\'ee au CNRS, UMR 8626}}\\
B\^at. 100, Universit\'e Paris-Sud, 91405 Orsay, France
\end{center}

\vskip 0.5cm
\centerline{\large Abstract}
\vskip 0.2cm

The thermodynamics of the anyon model projected 
on the lowest Landau level (LLL)
of an external magnetic field is
addressed in the anti-screening
regime, where the flux tubes carried by the anyons are parallel to the magnetic field.
It is claimed that the LLL-anyon equation of state, which is known in the screening regime, can be  analytically
continued in the statistical parameter across the Fermi point
to the antiscreening regime
up to the vicinity (whose width tends to zero
when the magnetic field becomes infinite) of the Bose point. There,
an unphysical  discontinuity arises due to the dropping of the 
non-LLL eigenstates which join the LLL, making the LLL
approximation  no longer valid. However, taking into account the effect of
the non-LLL states at the Bose point would only smoothen the discontinuity
and not alter the physics which is captured by the LLL projection: 
Close to the Bose point,
the critical filling factor  either goes to infinity (usual bosons)
in the  screening situation, or to 1/2 in the  anti-screening situation,
the difference between  the flux tubes orientation
being relevant even when they carry an infinitesimal fraction
of the flux quantum.
An exclusion statistics interpretation is adduced,
which explains this situation in  semiclassical terms. It
is further shown how the exact solutions of the 3-anyon problem support this
scenario as far as the third cluster coefficient is concerned.


\newpage
%\section[]{Introduction}

Identical particles with statistics continuously interpolating between
Bose-Einstein and Fermi-Dirac  exist in two \cite{Anyon}
and one \cite{Calogero-Sutherland} dimensions.
Contrary to the one-dimensional Calogero model,
which is solvable, the anyon spectrum is unknown.
However, a simplification arises when projecting the anyon model 
onto the lowest Landau level (LLL) of an external magnetic field,
 which is justified in the strong field/low temperature
limit.  A complete linear\footnote{Linear refers to the linear dependence in
$\alpha$ of the energy  in the presence of a long distance harmonic well
regulator.} eigenstate basis, which continuously interpolates
between the LLL-bosonic and the LLL-fermionic  basis, can be found
in the screening regime where the flux $\phi=\alpha\phi_o$
carried by the anyons  is antiparallel to the
external magnetic field---more precisely,
when the statistics parameter $\alpha$ 
which varies from  $\alpha=0$ (Bose) to
$\alpha=\pm 1$ (Fermi), is such that 
$\alpha\in [-1,0]$ if $eB>0$, or
equivalently $\alpha\in [0,1]$ if $eB<0$.

In this situation, the statistical mechanical properties  of 
the anyon gas have been derived \cite{Notre}.
Note that in the thermodynamic limit, both the LLL anyon and the Calogero 
models can be viewed as  microscopical realizations of
Haldane's exclusion statistics
\cite{Haldane,ES,Wu}. 
Various conformal field  theories have also been
shown to implement exclusion statistics \cite{SchoutensES-CFT}.

Clearly, since the magnetic
field gives a privileged orientation to the plane, one expects, for a given
magnetic field, quite different behavior depending on the
sign of $\alpha$, most particurlarly in the strong magnetic field limit.
Here, one addresses the question of the thermodynamics of the LLL
anyon model in the anti-screening
regime $\alpha\in [0,1]$ if $eB>0$ (or
$\alpha\in [-1,0]$ if $eB<0$), where  unknown nonlinear
 eigenstates should
become relevant when their gap above the LLL ground state vanishes in the
limit
$\alpha\to 0^+$ (or $\alpha\to 0^-$). 

Let us start with a short reminder on  the $N$-anyon model, 
which is defined
in the singular gauge by a  free $N$-body Pauli
Hamiltonian ($\hbar=m=1$)
$H_{\rm free}^{u}=-2\sum_{i=1}^N\partial_i\bar\partial_i
\quad,\quad H_{\rm free}^{d}=-2\sum_{i=1}^N\bar \partial_i\partial_i$,
where the index $u,d$ refers here to the spin degree of freedom.
The coupling to an external homogeneous magnetic field  amounts,
in the symmetric gauge, to
$\partial\to\partial-eB\bar z/4$ and
$\bar\partial\to\bar\partial+eBz/4$.
The $N$-body
eigenstates $\psi_{\rm free}$ of $H_{\rm free}$ have a nontrivial monodromy
encoded in the multivalued phase $ \exp(-i\alpha\sum_{k<
l}\theta_{kl})$
where $\sum_{k< l}\theta_{kl}$ is the
sum of the angles between pair relative radius vectors and
the $x$ axis in the plane. 
Looking at the multivalued phase as a singular gauge transformation,
one obtains, in the regular gauge,
an $N$-anyon  Aharonov-Bohm Hamiltonian
acting on single-valued wave functions  (bosonic by
convention) with contact interactions
$\mp\pi\alpha\sum_{i<j} \delta^2(z_i-z_j)$
(and $ \mp \sum_i eB/2$  energy shifts) induced by  the
spin up or spin down coupling to the local magnetic field of the
vortices (and to the external magnetic field). 
The contact interactions have to  implement
the exclusion of the diagonal of the
configuration space, and thus have to
be repulsive. So, depending
on the sign of $\alpha$, the spin up Hamiltonian  $(\alpha\in
[-1,0])$
or spin down Hamiltonian $(\alpha\in[0,1])$, is
used \cite{delta}.

Without loss of generality, let us
take in the sequel $eB>0$, i.e. $\omega_c=+eB/2$,
and ignore the trivial $\mp\sum_i eB/2$ Pauli induced shift.
Also, in order to compute  thermodynamic quantities,
a harmonic well of strength $\omega$ is added as a long distance regulator.
The thermodynamic limit will always be understood as $\omega\to 0$.

To encode in the eigenstates not only the anyonic multivalued phase
but also the short-range repulsion  and the long
distance Landau and harmonic exponential dampings
one  sets, if $\alpha\in [-1,0]$ (screening regime),
\be\label{0}
\psi_{\rm free}=\prod_{k<l}(z_k-z_l)^{-\alpha}\exp(-{1\over
2}\omega_t\sum_{i=1}^N z_i\bar z_i)\psi\ee
to obtain
an anyonic Hamiltonian acting on $\psi$ 
\bea\label{1}
{H} &=& -2  \sum _{i=1}^{N} 
              \left[ \partial_i\bar\partial_i 
                -{\omega_t+\omega_c\over 2}\bar z_i\bar \partial_i
		-{\omega_t- \omega_c\over 2} z_i \partial_i
		\right]
\nonumber \\
& & +2\alpha\sum_{i<j}\left[{1\over  z_i-
z_j}{(\bar \partial_i-\bar \partial_j})
-{\omega_t-\omega_c\over 2}\right]+\sum_{i=1}^N\omega_t
%\mp\omega_c
\eea
and, if $\alpha\in [0,1]$ (anti-screening regime),
\be\label{00} \psi_{\rm free}=
\prod_{k<l}(\bar z_k-\bar z_l)^{\alpha}\exp(-{1\over
2}\omega_t\sum_{i=1}^N z_i\bar z_i)\psi\ee
so that 
\bea\label{2}
{H} &=& -2  \sum _{i=1}^{N} 
              \left[ \partial_i\bar\partial_i 
                -{\omega_t+\omega_c\over 2}\bar z_i\bar \partial_i
		-{\omega_t- \omega_c\over 2} z_i \partial_i
		\right]
\nonumber \\
& & -2\alpha\sum_{i<j}\left[{1\over \bar z_i-
\bar z_j}{(\partial_i- \partial_j})
-{\omega_t+\omega_c\over 2}\right]+\sum_{i=1}^N\omega_t
%\mp\omega_c
\eea
with
$\omega_t = \sqrt{\omega_c^2+\omega^2}$.

Keeping in mind the strong magnetic field limit, one  
projects (\ref{1}) and (\ref{2}) on the LLL basis, that is,
on  $N$-body eigenstates made of symmetrized 
(i.e. with  bosonic quantum numbers 
$0\le\ell_1\le\ldots \le \ell_N $) products of 
the 1-body LLL holomorphic eigenstates 
\be \label{3}
({{\omega_c^{\ell_i+1}\over \pi \ell_i!}})^{{1\over 2}}z_i^{\ell_i},\quad
\ell_i\ge 0\ee
of energy $\omega_c$.

In the screening regime, the LLL projection of  the Hamiltonian (\ref{1}) 
in the thermodynamic limit gives, trivially, $H=N\omega_c$.
This is the LLL anyon model with an infinitely 
degenerate $N$-body spectrum. 
The virtue of the harmonic confinement
is  to lift the
degeneracy with respect to the $\ell_i$'s and
to bestow an explicit $\alpha$ dependence
on the $N$-body spectrum.
In a harmonic well, the  LLL eigenstates (\ref{3}) become 
\be \label{4}
({{\omega_t^{\ell_i+1}\over \pi \ell_i!}})^{{1\over 2}}z_i^{\ell_i},\quad
\ell_i\ge 0\ee
with a nondegenerate spectrum
\be\label{5}\omega_t+(\omega_t-\omega_c)\ell_i, \quad \ell_i\ge 0 \;.
\ee
Up to a $\omega_t$ dependent normalization,
the LLL anyonic eigenstates in a harmonic well rewrite as
\be\label{6}
\psi_{\rm free}=\prod_{i<j} (z_i-z_j)^{-\alpha}\prod_{i=1}^N z_i^{\ell_i}
\exp(-{1\over 2}\omega_t\sum_{i=1}^N z_i \bar z_i) \;.
\ee

Acting on the basis (\ref{4}), the Hamiltonian (\ref{1}) narrows down to
\be\label{7}
{H}_{\rm LLL} = N\omega_t+ \left[\sum_{i=1}^N
z_i \partial_i -{1\over 2} N(N-1)\alpha\right](\omega_t-\omega_c)
\ee
with a harmonic-LLL $N$-anyon spectrum
\be\label{8} E_N=N\omega_t+
\left[
\sum_{i=1}^N
\ell_i
-{1\over2}N(N-1)\alpha\right](\omega_t-\omega_c) \;.
\ee

The eigenstates and the spectrum
(\ref{6},\ref{8}) interpolate
from the harmonic-LLL bosonic to the
harmonic-LLL fermionic basis when $\alpha$ goes from 0 to $-1$
and lead, in the thermodynamic limit, to the
equation of state
\be\label{9} \beta P=\rho_L\ln(1+{\nu\over 1+\alpha\nu})\ee
and the  virial coefficients 
\be\label{10} a_n= (-{1\over \rho_{\rm L}})^{n-1} {1\over 
n}\{(1+\alpha)^n-\alpha^n\}\;.\ee

At the critical filling 
$\nu_{\rm cr}=-1/\alpha$
where the pressure diverges, one reaches  a nondegenerate ground state with all 
the $\ell_i$'s null: in the {\sl singular gauge}, it rewrites as 
\be\label{11} 
\psi_{\rm free}=\prod_{i<j} (z_{i}-z_{j})^{-\alpha}\exp(-{\omega_c\over 2} 
\sum_i^Nz_i\bar z_i) \;,
\ee
i.e., at the Fermi point $\alpha=-1$, 
the fermionic Vandermonde determinant
built from $1$-body Landau eigenstates.

Now the question is: What happens in the
anti-screening regime $\alpha\in [0,1]$?
The relevant Hamiltonian
(\ref{2}) no longer has a simple form  when it acts on products of
LLL holomorphic eigenstates (\ref{4}).
This translates into the fact that when $\alpha\to 0^+$
some, mostly unknown, non-LLL excited eigenstates 
join the ground state, as
indicated by various numerical and semi-classical analyses
\cite{Wu84,3annumerics,ChinHu,Dunne,Nanyon},
nd as can be appreciated explicitly in the solvable 2-anyon case:
The relative 2-anyon spectrum rewrites,
when the relative angular momentum $l$,
an even integer, satisfies $l\ge \alpha$, as
\be\label{14}  E_{\{n,l\}}=(2n+1)\omega_t
+(l-\alpha)(\omega_t-\omega_c)\to_{\omega\to 0}(2n+1)\omega_c\ee
\be\label{15} \psi_{{\rm free}\{n,l\}}=
z^{l-\alpha}L_n(\omega_t z\bar z)\exp(-{\omega_t\over 2} 
z\bar z)\ee
and when $l< \alpha$, as
\be\label{16} 
E_{\{n,l\}}=(2n+1)\omega_t+(\alpha-l)(\omega_t+\omega_c)\to_{\omega\to
0}(2n+1)\omega_c+2(\alpha-l)\omega_c\ee
\be\label{17} \psi_{{\rm free}\{n,l\}}=
\bar z^{\alpha-l}L_n(\omega_t z\bar z)\exp(-{\omega_t\over 2} 
z\bar z)\ee
with the wave functions
analytic (anti-analytic) in the relative coordinate 
$z=z_1-z_2$ for $l \ge \alpha$ ($l < \alpha$).

The bosonic LLL quantum numbers for the 2-body problem are
$n=0, l\ge 0$.
However, in the presence of the anyonic interaction, 
the LLL projection happens not to be  well defined at the bosonic
end $\alpha \to 0^+$.
Indeed, if $\alpha \in [-1,0]$, the LLL (analytic)
ground state basis obtained from (\ref{15}) by setting $n=0$ is
complete since the $l=0$ state belongs to this basis:
\be\label{18} E_{\{n=0,l\ge 0\}}=\omega_t
+(l-\alpha)(\omega_t-\omega_c)\ee
\be\label{19} \psi_{{\rm free}\{n=0,l\ge 0\}}=
z^{l-\alpha}\exp(-{\omega_t\over 2} 
z\bar z)\;.\ee
But if $\alpha\in [0,1]$, this same LLL
ground state basis becomes  incomplete
since the $l=0$ state is now anti-analytic 
[see (\ref{17})] with an  energy which varies linearly with $\alpha$,
joining the ground state basis when $\alpha \to 0^+$:
\be\label{20} E_{\{n=0,l\ge 2\}}=\omega_t
+(l-\alpha)(\omega_t-\omega_c)\ee
\be\label{21} \psi_{{\rm free}\{n=0,l\ge 2\}}=
z^{l-\alpha}\exp(-{\omega_t\over 2} 
z\bar z)\ee
and 
\be\label{22} E_{\{n=0,l=0\}}=\omega_t
+\alpha(\omega_t+\omega_c)\ee
\be\label{23}\psi_{{\rm free}\{n=0,l=0\}}=
\bar z^{\alpha}\exp(-{\omega_t\over 2} 
z\bar z)\;.\ee
It is easy to check that (\ref{21},\ref{23}) are eigenstates of the relative
part of the Hamiltonian   
(\ref{2}) taking into account the redefinition (\ref{00}).

\begin{figure}
\centerline{\hbox{
   \epsfig{figure=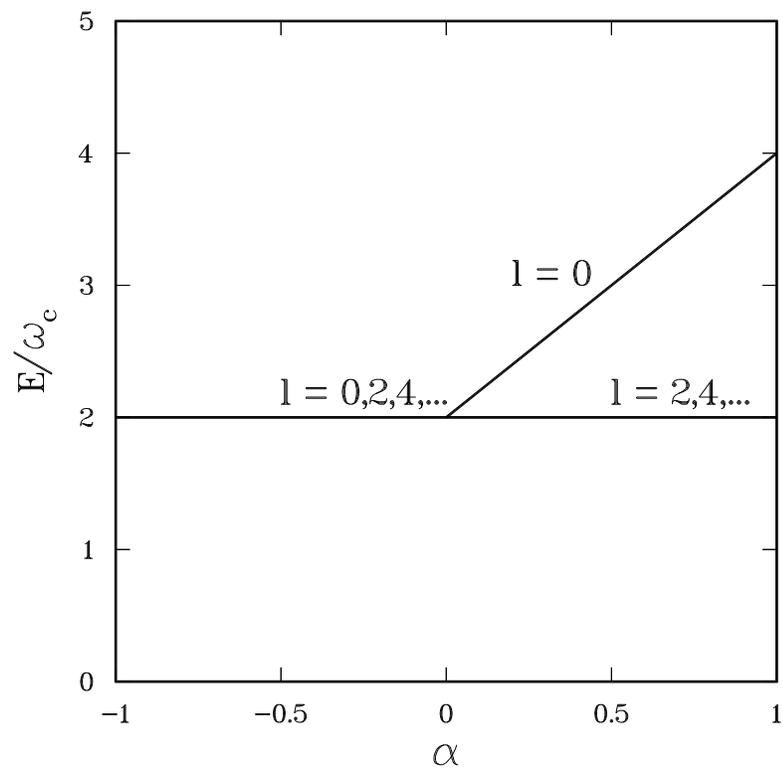,width=11cm}}
 }
 \caption{The lowest Landau level and the first excited state of the 2-anyon
 spectrum when $\omega=0$.}
 \label{phasediagram}
\end{figure}

What is the effect of the anti-analytic eigenstate on the $2$-anyon 
thermodynamics?
Consider the second virial coefficient \cite{2vcmagnetic}
 
%\be\label{24} a_2= \frac{\lambda^2}{x}
%               \bigg( -\frac{1}{4} \tanh x \, -\frac{1}{2}\alpha \,
%              -{{e^x (e^{-2\alpha x} -1)}\over
%             {4}}{( {{1}\over {\sinh x}} +{{1}\over{\cosh x}}) }
%             + (e^{x(|\alpha| -\alpha)}-1) \bigg) \quad\quad\quad\quad
%	   \ee
\be\label{24} a_2= \frac{\lambda^2}{x}
               \left[ -\frac{1}{4} \frac{1-e^{-2x}}{1+e^{-2x}} \,
               -\frac{1}{2}\alpha \,
              -\frac{e^{-2\alpha x} -1}{(1+e^{-2x})(1-e^{-2x})}
             + (e^{x(|\alpha| -\alpha)}-1) \right]
	   \ee
where $x=\beta\omega_c$
and  the thermal wavelength $\lambda=\sqrt{2\pi\beta}$.
\begin{figure}
\centerline{\hbox{
   \epsfig{figure=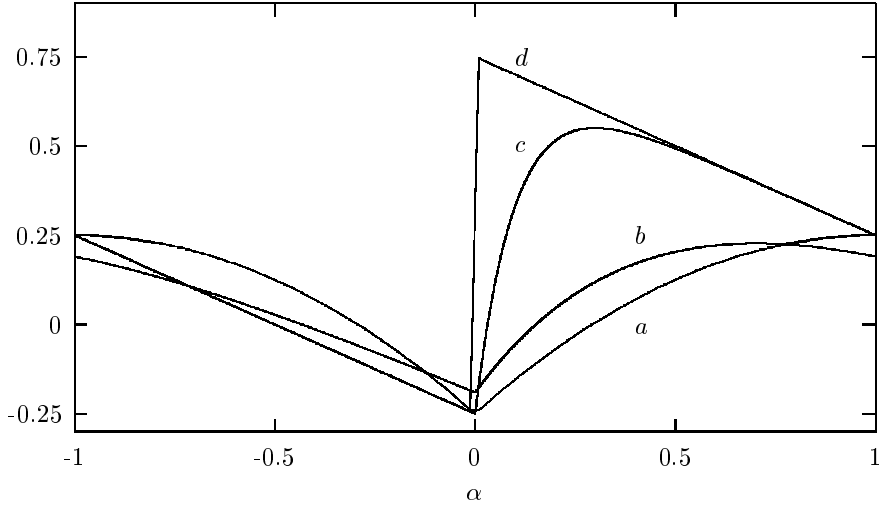,width=13cm}}
 }
 \caption{The second virial coefficient as a function of the
statistics parameter $\alpha$:   (a) $\frac{a_2}{\lambda^2}$ for zero 
magnetic field, 
(b) $\frac{a_2 \rho_L}{2}$ for $x=1$,   
(c) $ \frac{a_2 \rho_L}{2}$ for $x=5$,  
(d) $\frac{a_2 \rho_L}{2}$ for $x\to \infty$.}
 \label{v2ab}
\end{figure}
In the strong magnetic field  limit, $x\to\infty$,
\be\label{25}
a_2 = {1\over 2\rho_L}( -1 - 2 \alpha )
\qquad\mbox{for}\qquad \alpha\in[-1,0] \; , 
\ee
which indeed agrees with (\ref{10}), and
\be\label{26}
a_2 = {1\over 2\rho_L} \left( - 1 - 2 \alpha 
+ 4(1-e^{-2\alpha x}) \right) \qquad\mbox{for}\qquad \alpha\in[0,1] \; .
\ee

In considering the large $x$ behavior of (\ref{26}), the
order of limiting transitions, $x\to\infty$ and $\alpha \to 0^+$,
 is crucial. As long as $x$ is large but finite, $a_2$ is  a
continuous function of $\alpha$, since $1-e^{-2\alpha x}$
tends to zero as $\alpha \to 0^+$. However, if the $x\to \infty$
limit is taken first, (\ref{26}) becomes 
\be\label{28}
a_2 = {1\over 2\rho_L} \left( - 1 - 2 \alpha 
+ 4 \right) \qquad\mbox{for}\qquad \alpha\in\:]0,1] \;,  
\ee
so that $a_2$ is no longer continuous at $\alpha = 0^+$.
One can easily convince oneself that the discontinuity
is a direct consequence of dropping the $l=0$ eigenstate, since,
in the $x\to\infty$ limit, its energy gap with respect to
the ground state becomes infinite as soon as $\alpha\ne 0$.
In other words,
the effect of the excited state when $x$ becomes large is simply
to smoothen the
discontinuity, but not to alter the essence of the thermodynamics.
Its presence is only felt when $\alpha \sim 1/x$.

Note that (\ref{28}) is just (\ref{25}) with $\alpha-2$
substituted for $\alpha$. That is, one shifts $\alpha$ in (\ref{25}) by 2,
from $\alpha\in[-1,0]$ to $\alpha\in[1,2]$ (which is always legal,
because in the singular gauge one has  periodicity in $\alpha$
with period $2$) and one
continues it beyond the fermionic point $\alpha=1$,
where no peculiarity exists, down to $\alpha\in\:]0,1]$.
This is justified except near the
Bose point, in a vicinity whose width tends to zero with
the magnetic field tending to infinity, as discussed above.

In terms of the $2$-anyon  spectrum
(relative + center of mass), and leaving aside
the excited state, the analytic part of the
spectrum (\ref{18}-\ref{21}), i.e. the LLL spectrum,
rewrites when $\alpha\in [-1,0]$ as
\be\label{29} E_2=2\omega_t +(l_1+l_2-\alpha)(\omega_t-\omega_c)\ee
and  when $\alpha\in [0,1]$ as
\be\label{30} E_2=2\omega_t +(l_1+l_2+2-\alpha)(\omega_t-\omega_c)\ee
where  $0\le l_1\le l_2$ are bosonic quantum numbers. 
Note that computing $a_2$ directly from
the spectrum (\ref{29},\ref{30}) reproduces
(\ref{25},\ref{28}), and   when $\alpha\to 1$, (\ref{30}) rightly becomes 
the 2-fermion spectrum in the LLL 
 \be\label{31} E_2=2\omega_t +(l'_1+l'_2)(\omega_t-\omega_c)\ee
with  $0\le l'_1< l'_2$.

What we have just learned from the solvable
2-anyon system can be expected to be valid for the $N$-anyon system as 
well. Namely, starting from the exact $N$-anyon  spectrum (\ref{6},\ref{8}) 
when $\alpha\in
[-1,0]$, one can check that, by analogy with (\ref{29},\ref{30}),
 when $\alpha\in[0,1]$,
\be\label{32} E_N=N\omega_t+
\left[
\sum_{i=1}^N
\ell_i+{N(N-1)\over 2}(2
-\alpha)\right](\omega_t-\omega_c)\ee
and 
\be\label{33} \psi_{\rm free}=\prod_{i<j} (z_i-z_j)^{2-\alpha}\prod_{i=1}^N z_i^{\ell_i}
 \exp(-{1\over 2}\omega_t\sum_{i=1}^N z_i \bar z_i)
\ee
with $0\le l_1\le \ldots \le l_N$, are eigenvalues and eigenstates of
(\ref{2}) taking into account  the redefinition (\ref{00}).
When $\alpha\to 1$, both (\ref{32},\ref{33}) describe $N$ fermions with energy

\be\label{34}
 E_N=N\omega_t+(\omega_t-\omega_c)
\left[
\sum_{i=1}^N
\ell'_i\right]\ee
and  $0\le\ell'_1<\ldots < \ell'_N $.

We claim that (\ref{32},\ref{33})
captures the physics in the anti-screening regime, too, up to
the effect of the unknown eigenstates whose role, in the large $x$ limit, 
is only to smoothen the discontinuity in the
equation of state when $\alpha\to 0^+$.
The equation of state stemming from (\ref{32}) is
\be\label{35}  \beta P=\rho_L\ln(1+{\nu\over 1+(\alpha-2)\nu})\ee
with a critical filling 
$\nu_{\rm cr}=1/(2-\alpha)$
where the pressure diverges, describing a nondegenerate ground state with all 
the $\ell_i$'s null: in the {\sl singular gauge}, 
\be \label{36}
\psi_{\rm free}=\prod_{i<j}
(z_{i}-z_{j})^{2-\alpha}\exp(-{\omega_c\over 2} 
\sum_i^Nz_i\bar z_i) 
\ee
again becomes, when $\alpha=1$,  the fermionic  Vandermonde 
determinant built from $1$-body Landau eigenstates.

\begin{figure}
\centerline{\hbox{
   \epsfig{figure=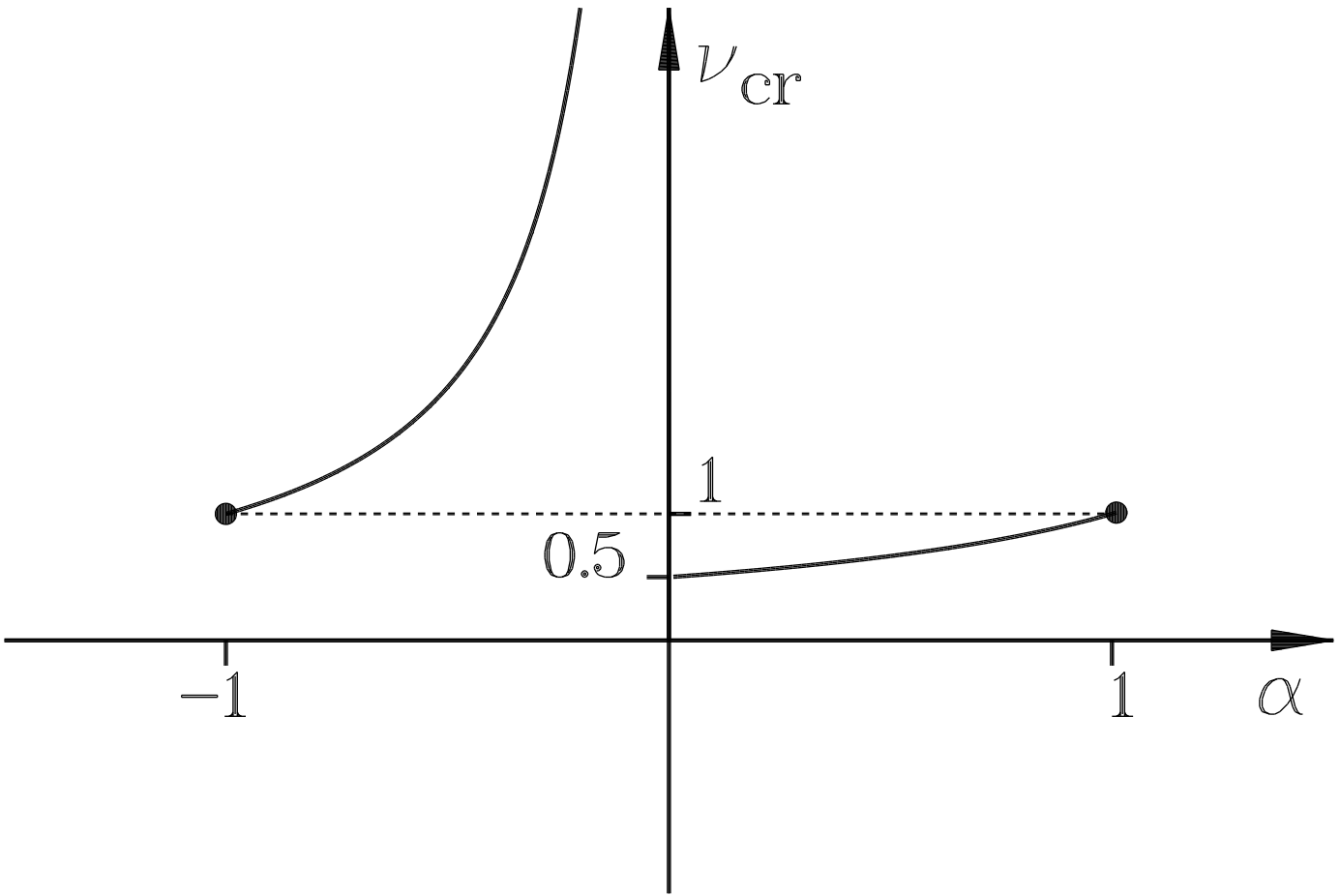}}
}
\caption{The critical filling  as a function of  $\alpha$.
There is a discontinuity at $\alpha=0^+$: the unknown nonlinear eigenstates
which join the groundstate at $\alpha=0^+$ smoothen the discontinuity.}
\label{}
\end{figure}

The physical situation in Fig. 3 is rather striking:
Moving away  from Bose statistics
by attaching infinitesimal flux tubes anti-parallel
to the magnetic field---the screening regime---leads to
a smooth interpolation between Bose ($\nu_{\rm cr}=\infty$) and Fermi 
($\nu_{\rm cr}=1$) statistics,
whereas attaching
infinitesimal statistical flux tubes
parallel to the magnetic field---the anti-screening regime---condenses the 
system into a $\nu_{\rm cr}=1/2$ quantum  system.

The equation of state taking the same form both in the screening and
antiscreening regime, 
so should do its interpretation in terms of      
exclusion statistics. It is possible to adduce
a simple semiclassical picture, within the
approach of Ref.~\cite{IM}, that helps understand the
metamorphosis with the exclusion statistics parameter
and, by consequence, the filling factor.
One starts with semiclassical
single-particle orbits in the harmonic-Landau potential,
which are formed by two normal modes with frequencies
$\omega_\pm = \pm\omega_t - \omega_c$. The ``splitting''
($\omega_+$) mode
corresponds to the lifted degeneracy of the Landau levels,
in particular of the LLL.
Note that $\omega_+\to 0$ when $\omega\to 0$.
The Landau ($\omega_-$) mode corresponds to different
Landau levels; thus, within the LLL it cannot be excited.
The corresponding orbits are circles with opposite directions of
rotation for the two modes, as evidenced by the opposite signs of
$\omega_+$ and $\omega_-$. Single-particle LLL orbits, without regard
for statistical interaction, are concentric circles, with the value $l$
of the angular momentum, an integer, corresponding to the number
of quanta of the excited splitting mode. Bosons can all be in
the ground state $l=0$, fermions must occupy the $l=0,1,\ldots$ states,
one particle per state. Now consider two anyons
in this picture. The first one can sit in the ground state,
and, the second one being in an excited state with angular momentum
$l$, their statistical exchange phase will be $\exp(i\pi l)$.
Demanding this to be equal to $\exp(-i\pi\alpha)$ yields
$l=-\alpha,\;2-\alpha,\;4-\alpha,\;\ldots\;$. For $\alpha$ negative, the lowest
allowed angular momentum of the second particle is thus $l=-\alpha$,
which corresponds to exclusion statistics with statistics parameter
$g=-\alpha$ (the presence of the first particle excludes $g$ states,
in terms of the values of the quantum number, for the second one).
However, for $\alpha$ positive, the would-be lowest value of $l$
is negative, which is prohibited: it corresponds to the opposite
direction of rotation, that is, to an excitation of the
Landau mode. This is, in a different language, precisely
the same thing as the two-particle ground state detaching from
the LLL basis [cf. (\ref{22})]. The true lowest allowed value of $l$
belonging to the splitting mode for $\alpha>0$
is therefore $2-\alpha$, and the same is the exclusion statistics
parameter: $g=2-\alpha$. This corresponds exactly to the equation
of state (\ref{35}). Imposing the condition that
the Landau mode may not be excited at all corresponds to
taking the $\omega_c\to\infty$ limit first; the whole construction
is then valid for any $\alpha\in[0,1]$.

To deduce the complete spectrum, one still has, in fact,
to refer to the underlying quantum-mechanical problem.
In particular, the semiclassical picture alone will not explain
why $l=3-\alpha$ is allowed for the second particle (it has to be
in order to get the count of excited states right) while
$l=1-\alpha$ is not; the answer is that quantum-mechanically, those  
contain the $l_\mathrm{CM} = 1$ center-of-mass excitations
over, respectively, $l_\mathrm{rel} = 2-\alpha$ (which
is allowed) and $l_\mathrm{rel} = -\alpha$ (which is not).
The inherent problem with the semiclassical picture per se is that
single-particle
angular momenta are not good quantum numbers for anyons.
However, with the quantum-mechanical knowledge put in
as outlined, a generalization to $N$ particles is possible
and yields the exact result in the LLL anti-screening
regime, like it does in the screening regime \cite{IM}.
Semiclassically, the third particle has the first ($l=0$)
and second ($l=2-\alpha$) ones inside its orbit
and therefore has to have its angular momentum equal 
to $n-2\alpha$, to provide for the correct Aharonov-Bohm phase.
Now, $n=3$ is excluded because quantum-mechanically, that
would involve as one of the states in the superposition the
state with the relative angular momentum of the second and
third particles equal to $-\alpha$, which is not allowed
(Landau excitation), leaving $4-2\alpha$ as the lowest
possible angular momentum for the third particle.
Continuing to $N$ particles, the ground state energy
$E_{0N} = \frac{1}{2}N(N-1)(2-\alpha)(\omega_t-\omega_c)$
is thus correctly reproduced, up to the constant shift.
The crucial difference in the critical filling factor
between the two directions of the infinitesimal flux
is thereby interpreted in terms of exclusion statistics:
With screening, adding a particle excludes only $\alpha$
quantum states because the sign of the angular momentum
remains the one that belongs to the LLL, but with 
anti-screening, the sign gets reversed and the would-be
ground state is promoted to the next Landau level, which
is why the whole of $2-\alpha$ (the next allowed value
of the relative angular momentum) states get excluded.

By way of supporting this claim quantitatively,
it makes sense, as a first step,  to look at the $3$-anyon problem
and its thermodynamics,  described by 
the third virial coefficient $a_3$, or by
the third cluster  coefficient $b_3$ from which $a_3$ is deduced.
One has linear eigenstates
\cite{Wu84,Dunne}, which generalize the linear eigenstates (\ref{15},\ref{17})
of the 2-body problem, and nonlinear
eigenstates, which are only known numerically \cite{3annumerics}.
In the absence of a magnetic field, the latter have been shown
\cite{3anvirial} to render $a_3$ finite and continuous for
all values of $\alpha$. In the strong magnetic field limit,
however, one would expect $a_3$  to
be mainly controlled  by the linear LLL eigenstates,
since the (mostly nonlinear)
non-LLL eigenstates are excited and  exhibit a gap proportional
to $x$. Put it differently, ignoring these  states should
amount to no more than excluding the $l=0$ excited state
in the 2-anyon problem, and therefore introducing 
an unphysical   discontinuity of the equation of state in
the vicinity (of width $\sim 1/x$) of the bosonic point, $\alpha=0^+$,
where some nonlinear and linear states join the ground state.

Restricting oneself to the contribution of the linear states,
one  obtains for $\alpha\in [0,2]$
\bea\label{37}{\nonumber} b_3^{\rm lin}
&=& {e^{-3(\tilde {x}+x)}\over (1-e^{-\tilde
{x}})( 1-e^{-\tilde
{x}-2x})}\left[{{e^{-3\tilde {x}(2-\alpha)}
     \over (1-e^{-2\tilde
{x}})(1-e^{-3\tilde
{x}})(1-e^{-2\tilde
{x}-2x})(1-e^{-3\tilde
{x}-2x})}}\right.\\
{\nonumber} &-& \left.
{e^{-\tilde {x}(2-\alpha)}
     \over( 1-e^{-2\tilde{x}})(1-e^{-\tilde{x}})
( 1-e^{-2\tilde{x}-2x})(1-e^{-\tilde{x}-2x})}
+{1\over 3(1-e^{-\tilde{x}})^2(1-e^{-\tilde{x}-2x})^2}
\right]\\
{\nonumber} &+&
{e^{3(\tilde {x}+2x)(2-\alpha)}e^{3(\tilde {x}+x)}
\over    ( 1-e^{\tilde
{x}})(1-e^{\tilde
{x}+2x})( 1-e^{2\tilde
{x}+4x})(1-e^{3\tilde
{x}+6x})( 1-e^{2\tilde
{x}+2x})(1-e^{3\tilde
{x}+4x})}\\
&-&
{e^{(\tilde {x}+2x)(2-\alpha)}e^{\tilde {x}+x}
\over    ( 1-e^{\tilde
{x}})(1-e^{\tilde
{x}+2x})( 1-e^{2\tilde
{x}+4x})(1-e^{2\tilde
{x}+2x})( 1-e^{-\tilde
{x}})(1-e^{-\tilde
{x}-2x})}
\eea
where $\tilde{x}=\beta(\omega_t-\omega_c)$.
Taking  in (\ref{37}) the thermodynamic limit, whence
$\tilde {x}\to(\beta\omega)^2/2x$, 
%b_3={1\over \tilde{x}^2}{1\over 1-e^{-4x}}{e^{-x}e^{2(2-\alpha)x}\over
%(1-e^{2x})^3}{1\over \tilde{x}}()+\ldots 
one sees that $b_3^{\rm lin}$ exhibits an unphysical volume divergence at leading order
$1/{\tilde {x}} ^2$ which is obviously due to the dropping of the  infinite set 
of unknown nonlinear eigenstates.
In the LLL limit
$x\to\infty$, (\ref{37}) becomes
\be b_3^{\rm lin}\simeq 
e^{-3x}\left[-{1\over {\tilde{x}}^2}e^{-2\alpha x}+{1\over
18{\tilde{x}}}(9(2-\alpha)^2-9(2-\alpha)(1+2e^{-2\alpha x})+2(1+9e^{-6\alpha
x}+36e^{-2\alpha x})\right]
\ee
Again, the unphysical $1/{\tilde {x}} ^2$ volume divergence would not appear
if the nonlinear states were included.  We are interested in the term with
the $1/\tilde x$ volume divergence
which makes, in the thermodynamic limit, the cluster coefficient
proportional to the volume, as it should be. When this term is
periodically extended from $\alpha\in[1,2]$ onto $\alpha\in[-1,0]$,
a discontinuity similar to the one
described in the $2$-anyon case arises. Nicely enough, both
the unphysical volume divergence and the LLL discontinuity
are controlled  by the very same exponential factor $e^{-2\alpha x}$ which was
already operative in the 2-anyon problem.
Putting all these considerations together, we obtain, in the thermodynamic limit \cite{regularization} $1/{\tilde
{x}}\to 3\rho_LV$ and in the large magnetic field limit,
for $\alpha\in [-1,0]$ 
\be  b_3=\rho_l V e^{-3x}{(3\alpha+1)(3\alpha+2)\over 3!}\ee
which is consistent with  (\ref{10}),
and for $\alpha\in\:]0,1]$
\be b_3=\rho_l V e^{-3x}{(3(\alpha-2)+1)(3(\alpha-2)+2)\over 3!}\ee
which is consistent with (\ref{32}) and (\ref{35}).
Again, the effect of the  linear and nonlinear states
joining the LLL at $\alpha=0^+$ would amount to smoothening this
discontinuity at the Bose point. Clearly, a generalization
of these results to the $N$-anyon case should follow the same lines of
reasoning.

To conclude, let us recapture our main claim again.
The LLL anyon equation of state is continuous through the
Fermi point but behaves in two much different ways near the Bose
point, depending on which side the latter is approached from.
In the screening regime, it tends to the bosonic equation
in a smooth manner, with the critical filling factor
going to infinity as $\alpha\to0^-$. In the anti-screening
regime, however,
there is an abrupt change on a narrow
interval near the Bose point due to extra states joining
the LLL at $\alpha=0^+$. If the $B\to\infty$ limit is
taken first, that is, if one ignores these extra states, an unphysical 
discontinuity arises,
and the critical filling factor at $\alpha\to0^+$
then tends to $1/2$.
In reality, at no matter how small positive $\alpha$,
the critical filling  tends to $1/(2-\alpha)$ as $B\to\infty$.
It  being believed that the $\nu_{\rm cr}=1/2$ state
should play an important role in the fractional quantum
Hall effect (see for example the composite fermion approach and the
resulting Jain series \cite{Jain}),
the LLL-anyon model provides a scenario
of how the $1/2$ filling may arise without relying on extra
interaction (like the Coulomb interaction that plays a
crucial role in the usual picture of FQH states), but
just from the interplay of the strong
magnetic field and statistics close to the Bose point in the
anti-screening regime.

S.O. would like to thank G. Lozano and S. Isakov 
for discussions and early collaboration on the subject. We acknowledge 
numerous useful discussions with J. Myrheim and K. Olaussen.

\end{document}